\documentclass[12pt,preprint]{aastex}

\pdfoutput=1

\newcommand{\HDblah}{HD~$\!$209458~$\!$b}
\def\gsim{\;\rlap{\lower 2.5pt\hbox{$\sim$}}\raise 1.5pt\hbox{$>$}\;}
\def\lsim{\;\rlap{\lower 2.5pt\hbox{$\sim$}}\raise 1.5pt\hbox{$<$}\;}
\def\del{{\partial}}

\begin{document}

\title{Atmospheric Circulation of Hot Jupiters: A Shallow
Three-Dimensional Model}

\author{Kristen Menou \& Emily Rauscher}

\affil{Department of Astronomy, Columbia University,\\ 550 W. 120th
  Street, New York, NY 10027, USA}

\begin{abstract}
Remote observing of exoplanetary atmospheres is now possible, offering
us access to circulation regimes unlike any of the familiar Solar
System cases.  Atmospheric circulation models are being developed to
study these new regimes but model validations and intercomparisons are
needed to establish their consistency and accuracy. To this end, we
present a simple Earth-like validation of the pseudo-spectral solver
of meteorological equations called IGCM (Intermediate General
Circulation Model), based on Newtonian relaxation to a prescribed
latitudinal profile of equilibrium temperatures. We then describe a
straightforward and idealized model extension to the atmospheric flow
on a hot Jupiter with the same IGCM solver. This shallow,
three-dimensional hot Jupiter model is based on Newtonian relaxation
to a permanent day-night pattern of equilibrium temperatures and the
absence of surface drag. The baroclinic regime\footnote{Contours of
pressure and density are misaligned in a baroclinic flow, while they
are aligned in a barotropic flow. Strongly baroclinic flows are
susceptible to baroclinic instabilities which, in essence, are slanted
versions of convection \citep[see the review by][]{Showman2007}.} of
the Earth's lower atmosphere is contrasted with the more barotropic
regime of the simulated hot Jupiter flow. For plausible conditions at
the 0.1-1 bar pressure level on HD 209458b, the simulated flow is
characterized by unsteadiness, subsonic wind speeds, a
zonally-perturbed superrotating equatorial jet and large scale polar
vortices. Violation of the Rayleigh-Kuo inflexion point criterion on
the flanks of the accelerating equatorial jet indicates that
barotropic (horizontal shear) instabilities may be important dynamical
features of the simulated flow. Similarities and differences with
previously published simulated hot Jupiter flows are briefly noted.
\end{abstract}


\section{Introduction}

Since the first discoveries of extrasolar giant planets around nearby
sun-like stars with Doppler velocimetry (Mayor \& Queloz 1995; Marcy
\& Butler 1996), extrasolar planet research has experienced a
spectacular series of observational breakthroughs. In recent years,
progress has been particularly rapid with a subset of close-in
extrasolar planets found to transit the disk of their host star
\citep[e.g.,][]{Charbo2008,Deming2008}.

The first transit observation of a hot Jupiter (Charbonneau et al.\
2000; Henry et al.\ 2000) has been followed by many such measurements,
including a growing number of new close-in planet discoveries based on
transit searches \citep[e.g.,][]{Torres2008}. For the brightest nearby
systems with transiting planets, this has also enabled the detection
of the planetary thermal flux occulted at secondary eclipse (e.g.,
Deming et al. 2005, 2006; Charbonneau et al.  2005; Harrington et
al. 2007), the detection of day/night temperature variations through
IR phase curve monitoring (e.g. Harrington et al. 2006; Knutson et
al. 2007, 2008b; Cowan et al. 2007), IR spectral measurements
(Grillmair et al. 2007; Richardson et al. 2007; Knutson et al. 2008a)
as well as a variety of additional constraints based on transit
spectroscopic studies (Tinetti et al. 2007; Ehrenreich et al. 2007;
Swain et al. 2008; Barman 2007; Redfield et al. 2008; Pont et
al. 2008).  Overall, these results have contributed to a shift in
focus from the detection of extrasolar planets to the characterization
of their physical atmospheric properties
\citep[e.g.,][]{Seager2005,Burrows06,Marley2007,Fortney2007,Barman2008,Burrows2008}.

The need to interpret these astronomical data reliably, so as to infer
the physical conditions present in distant exoplanetary atmospheres,
has also fueled a growing atmospheric modeling effort.  While
plane-parallel radiative models have been the tools of choice to
interpret these data until now, they are likely insufficient in the
case of hot Jupiter/Neptune atmospheres.  Indeed, due to their short
orbital separations, all these exoplanets are expected be
tidally-locked to their parent star (or, in some cases,
pseudo-synchronized). Tidally-locked planets are subject to an unusual
situation of permanent, asymmetric day/night radiative forcing,
leading to heat redistribution by atmospheric motions.  Current data
on hot Jupiters already indicate that existing one-dimensional
radiative transfer models fail to capture the multi-dimensional nature
of this atmospheric regime (e.g., Seager et al. 2005; Knutson et
al. 2007; 2008b; Fortney et al. 2006) and, when included, the effects
of atmospheric heat redistribution are usually accounted for only
indirectly
\citep[e.g.,][]{Iro05,Seager2005,Barman2008,Burrows2008,Fortney2008}. To
adequately address this circulation regime and interpret the growing
data set, multi-dimensional, coupled radiation-hydrodynamics models of
these atmospheres are required \citep[see the review
by][]{Showman2007}.

Hot Jupiters have high atmospheric temperatures, slow rotation rates
and an unusually permanent pattern of asymmetric day/night radiative
forcing. Recognizing that this constitutes an interesting problem in
atmospheric dynamics, several groups have explored plausible
circulation regimes on these planets using different modeling
approaches and assumptions
\citep{Showman2002,Cho2003,Menou2003,C&S05,Cooper2006,Langton2007,Cho2006,Dobbs-Dixon2008,Langton2008,Showmanetal08}. These
investigations have exhibited significant diversity in flow results,
for which there currently is no simple unifying explanation \citep[see
the discussion in][]{Showman2007}.

Atmospheres in motion are non-linear, presumably turbulent flows. In
addition, radiatively active species in an atmosphere are advected by
the flow and in doing so non-linearly couple the flow to its effective
source of radiative forcing. Simulated atmospheric flows are thus
generally expected to be sensitive to various numerical and physical
details of any specific implementation.  In the exoplanetary context,
with rather limited direct information on the atmospheres studied,
this means that simple parameterized models isolating key features of
the simulated flow are important in helping us understand the general
behavior of remote atmospheric flows. With this in mind, we present
here idealized atmospheric models which emphasize dynamical aspects
under linear forcing conditions. These simplified models do not
address any detailed aspect of the radiative or chemical structure of
the atmospheres under consideration.

In general, different numerical implementations of a specific
atmospheric problem will not lead to identical results. For this
reason, model validations and inter-comparisons are standard practices
in atmospheric science. This is particularly true of the complex
hydrodynamic solvers known as dynamical cores
\citep[e.g.,][]{HeldSuar94}.  The development of dynamical cores to
solve the equations satisfied by an hydrostatic atmosphere in motion
has been a major enterprise. Dynamical cores currently used in
meteorological and climate models are the results of years of
refinements to guarantee stability, efficiency, and accurate
conservation of mass, momentum and energy
\citep[e.g.,][]{HoskSimm75}. In the exoplanetary context, different
modeling groups have used various hydrodynamic solvers. Some are new
and little tested while others are old, well-tested but only validated
under conditions appropriate for Solar System planetary
atmospheres. As a result, intercomparisons and validations of
dynamical cores on idealized atmospheric problems tailored for
exoplanets will be important in the future to assess, both
qualitatively and quantitatively, how reliable our interpretations of
exoplanetary data can be.

As a first step in this direction, we present here a simple Earth-like
validation of the IGCM dynamical core (in \S3), followed by a direct
extension of this model to the atmospheric flow on a hot Jupiter (in
\S4).

\section{The IGCM Dynamical Core}

The IGCM dynamical core has its origins in the development of accurate
numerical solvers for meteorological studies \citep{HoskSimm75}. It is
a well tested and documented solver, which has been used for studies
as diverse as the climate of the Earth \citep{deFor2000}, the
meteorology of Mars \citep{Joshi95, Collins95} and the atmospheric
circulation regime of tidally-locked terrestrial exoplanets
\citep{Joshi97}. We summarize the main features of this dynamical
solver here.

\subsection{Equations of Motion}

We consider a coordinate system rotating with the same angular
velocity, $\Omega_p$, as the planet. Longitude $\lambda$ ranges from
$0$ to $2 \pi$ in the eastward direction and latitude $\phi$ ranges
from $- \pi /2$ at the south pole to $+ \pi /2$ at the north
pole. From the general fluid equations satisfied by an inviscid ideal
gas, one derives the primitive equations of meteorology in the
traditional approximation, which are satisfied on large horizontal
scales\footnote{The primitive equations cease to be satisfied by
motions with horizontal scales approaching the atmospheric vertical
scale height (e.g., Holton 1992; Vallis 2006).}  by a hydrostatic
atmosphere in motion. Using pressure, $p$, as the vertical coordinate,
equations for the conservation of momentum, mass and energy can be
written, respectively,
\begin{mathletters}\label{primitive equations}
\begin{eqnarray}
\frac{d{\bf v }}{d t} + 
\frac{u\tan\phi}{R_p}\, {\bf k}\!\times\!{\bf v}\! &\! =\! &\!
-\nabla\! _p\ \Phi - f{\bf k}\!\times\!{\bf v} ,\\
\frac{\del\Phi}{\del p}\! &\!  =\! &\! -\frac{1}{\rho} \,\\
\frac{\del\omega}{\del p} &\! =\! &\! -\nabla\! _p\cdot{\bf v}\ ,\\ 
c_p \frac{d T}{d t} &\! =\! &\! \frac{\omega}{\rho}\ + Q\ ,
\end{eqnarray}
\mbox{where the Lagrangian derivative is}
\begin{equation}
\frac{d}{d t} \ = \ \frac{\del}{\del t} + 
{\bf v}\!\cdot\!\nabla\! _p + \omega\!\frac{\del}{\del p},\ 
\end{equation}
\end{mathletters}\noindent
${\bf v}\! =\! (u,v)$ is the (eastward, northward) horizontal velocity
in the rotating frame, $R_p$ is the planetary radius and $\omega
\equiv dp/dt$ is the vertical ``velocity.'' The geopotential $\Phi =
gz$, where $z$ is the height above a fiducial pressure surface and $g$
is the gravitational acceleration, which is assumed to be constant and
includes the contribution from the centrifugal acceleration. The unit
vector ${\bf k}$ is normal to the ``planetary surface.'' The Coriolis
parameter $f = 2 \Omega_p \sin \phi$ is the projection of the
planetary vorticity vector, $2{\bf\Omega}_p$, onto ${\bf k}$. The
operator $\nabla\!  _p$ is the horizontal gradient on a constant-$p$
surface. In the energy equation, $c_p$ is the heat capacity of the
atmospheric gas at constant pressure and $Q$ is the diabatic
heating/cooling rate. The above equations are closed with the ideal
gas law, $p = \rho {\cal R}T$, as the equation of state, where ${\cal
R}$ is the atmospheric perfect gas constant. The validity of the
primitive equations in the traditional approximation for the study of
hot Jupiter atmospheric flows has been discussed, for instance, by
\citet{Showman2007} and \citet{Cho2006}.

The IGCM dynamical core solves these equations in a form that involves
the vorticity and divergence of the horizontal flow. Using scaled
units of length ($\propto R_p$), time\footnote{All times are
subsequently quoted in units of a planetary day equal to $2 \pi /
\Omega_p$.} ($\propto \Omega_p^{-1}$) and temperature ($\propto R_p^2
\Omega_p^2 / {\cal R}$), as well as rescaled zonal and meridional
velocities
\begin{eqnarray} 
U & = &  u \cos \phi ,  \nonumber \\
V  & = & v \cos \phi ,  \nonumber
\end{eqnarray}
the horizontal momentum equations specifically solved are the equation
for the vertical component of absolute vorticity, $\zeta$,
\begin{equation} \label{eq:vor}
\frac{\partial \zeta}{\partial t} = \frac{1}{1- \mu^2} \frac{\partial
{\cal F}_V}{\partial \lambda} - \frac{\del {\cal F}_U}{ \del \mu} +
Q_{\rm vor} + Q^{\rm hyp}_{\rm vor},
\end{equation}
 and the equation for the horizontal divergence, $D$,
\begin{equation} \label{eq:div}
\frac{\partial D}{\partial t} = \frac{1}{1- \mu^2} \frac{\partial
{\cal F}_U}{\partial \lambda} + \frac{\del {\cal F}_V}{ \del \mu} -
\nabla^2 \left[\frac{U^2 + V^2}{2 (1 - \mu^2)} + \Phi + T_{\rm ref}
\ln p_{\rm surf} \right] + Q_{\rm div} + Q^{\rm hyp}_{\rm div},
\end{equation}
 where
\begin{eqnarray} 
\mu  &\equiv& \sin \phi \nonumber \\
\zeta &\equiv& 2 \mu+ \frac{1}{R_p \cos \phi}\frac{\del v}{\del \lambda} -  \frac{1}{R_p \cos \phi} \frac{\del \,u cos \phi}{\del \phi} \nonumber \\
D &\equiv& \frac{1}{R_p \cos \phi}\frac{\del u}{\del \lambda} +  \frac{1}{R_p \cos \phi} \frac{\del \, v cos \phi}{\del \phi} \nonumber \\
{\cal F}_U & = & V \zeta - \dot \sigma \frac{\del U}{\del \sigma} -
T_{\rm a} \frac{\del \ln p_{\rm surf}}{\del \lambda} \nonumber \\ 
{\cal F}_V & = & -U
\zeta - \dot \sigma \frac{\del V}{\del \sigma} - T_{\rm a}(1-\mu^2)
\frac{\del \ln p_{\rm surf}}{\del \mu}.\nonumber 
\end{eqnarray}
Conservation of energy is expressed via the temperature equation
\begin{equation} \label{eq:temp}
\frac{\partial T_{\rm a} }{\partial t} = - \frac{1}{1- \mu^2}
\frac{\partial}{\partial \lambda} U T_{\rm a}-\frac{\del} { \del \mu} V
T_{\rm a} +D T_{\rm a}- \dot \sigma \frac{\del T}{\del \sigma} +
\kappa \frac{T \omega}{p} + Q_T + Q^{\rm hyp}_T,
\end{equation}
where the temperature $T =T_{\rm ref} + T_{\rm a}$ is decomposed into
an arbitrary (constant) reference value, $T_{\rm ref} (\sigma)$,
around which the anomaly, $T_{\rm a} (\sigma)$, is
calculated. Conservation of mass is expressed via an equation for the
vertically integrated ``surface'' pressure,
\begin{equation}
\frac{\del \ln p_{\rm surf}}{\del t}= - \frac{U}{1- \mu^2} \frac{\partial \ln
p_{\rm surf} }{\partial \lambda} - V \frac{\del \ln p_{\rm surf}}{\del
\mu} - D - \frac{\del \dot \sigma}{\del \sigma},
\end{equation}
where $p_{\rm surf}$ is the pressure at the bottom model level, due to
the entire weight of the overlaying atmosphere. Momentum balance in
the vertical satisfies the same hydrostatic equation as before,
\begin{equation} \label{eq:hydsta}
\frac{\del \Phi}{\del \ln \sigma} = - T.
\end{equation}
In all the above equations, the vertical pressure coordinate $\sigma =
p/p_{\rm surf}$ is used ($0 \leq \sigma \leq 1$).  These five
equations are the exact same ones as originally solved by
\citet{HoskSimm75}, except for the various additional $Q$ terms
entering the vorticity, divergence and temperature equations.

\subsection{Forcing, Friction and Hyperdissipation}

Our IGCM implementation contains linear prescriptions for temperature
relaxation as well as drag on the vorticity and divergence
fields. Newtonian relaxation of the temperature field to a prescribed
equilibrium temperature profile, $T_{\rm eq}$, on a characteristic
radiative time $\tau_{\rm rad}$, corresponds to
\begin{equation}
Q_T = \frac{T_{\rm eq} - T}{\tau_{\rm rad}}
\end{equation}
in equation~(\ref{eq:temp}). This linear temperature relaxation scheme
is a very simplified form of atmospheric forcing which encapsulates
all the unspecified chemical-radiative physics in the modeled
atmosphere.  Rayleigh drag on the horizontal flow field, on a
characteristic friction time $\tau_{\rm fric}$, corresponds to
\begin{eqnarray} 
Q_{\rm vor} & = & - \frac{\zeta - 2 \mu}{\tau_{\rm fric}} \\ 
Q_{\rm div} & = &- \frac{D}{\tau_{\rm fric}}
\end{eqnarray}
in equations~(\ref{eq:vor}) and~(\ref{eq:div}), respectively. (Note
that friction is only applied to the relative vorticity field.)

In this work, Rayleigh drag is used as a simple parameterization of
surface friction in the Earth-like model only, while Newtonian
relaxation is used as the unique form of forcing in all models. The
vertical profile of equilibrium temperature used has two regions, a
lower atmosphere (troposphere) where temperature decreases with height
at a fixed rate, and an upper atmosphere (stratosphere) which is
vertically isothermal. The specific relaxation temperature profile
adopted in all the models has the form
\begin{equation} \label{eq:teqz} 
T^{\rm vert}_{\rm eq}(z) = T_{\rm surf} - \Gamma_{\rm trop} (z_{\rm stra} +
\frac{z- z_{\rm stra}}{2}) + \sqrt{ \left( \frac{1}{2} \Gamma_{\rm
trop} [z- z_{\rm stra}] \right)^2+ \delta T_{\rm stra}^2 },
\end{equation}
where $z$ is the height above the lowest ``surface'' level, $T_{\rm
surf}$ is the ``surface'' temperature, $\Gamma_{\rm trop}$ is the
lapse rate ($\equiv - dT/dz$ ) in the tropospheric region, $z_{\rm
stra}$ is the height at which one enters the stratospheric region and
$\delta T_{\rm stra}$ is a temperature offset used to smooth out the
transition between the finite tropospheric lapse rate and the
isothermal stratosphere.

Differential heating and cooling of atmospheric regions around the
planet drives atmospheric motions. In addition to the vertical
dependence in equation~(\ref{eq:teqz}), one must also specify the
latitudinal and longitudinal dependence of the temperature relaxation
profile. In our models, the three-dimensional equilibrium temperature
is given by
\begin{equation} \label{eq:Teq}
T_{\rm eq}(\sigma, \lambda, \phi) = T^{\rm vert}_{\rm eq}(\sigma) +
\beta_{\rm trop}(\sigma) \Delta T_\theta (\lambda,\phi),
\end{equation} where
\begin{equation} \label{eq:Teq2}
 \beta_{\rm trop}(\sigma) =  \sin \frac{\pi (\sigma - \sigma_{\rm stra})}
{2(1-\sigma_{\rm stra})} \nonumber
\end{equation}
is a height-dependent damping factor applied in the troposphere
($\sigma \geq \sigma [z_{\rm stra}]$) to gradually reduce the
latitudinal and longitudinal temperature differential over the
vertical extent of the troposphere. In the stratosphere, this
horizontal temperature differential is set to zero: $ \beta_{\rm
trop}(\sigma < \sigma [z_{\rm stra}]) = 0$.  The correspondence
between $z$ and $\sigma$ levels is obtained by vertical integration of
the hydrostatic balance equation (Eq.~[\ref{eq:hydsta}]). The
latitudinal and longitudinal dependence of the relaxation temperature
profiles, $\Delta T_\theta (\lambda,\phi)$, are specified separately
for the Earth-like and hot Jupiter models in \S~\ref{sec:Earth}
and \S~\ref{sec:HJ} below.

When using the primitive equations, one assumes that effects on small
scales, which are not adequately described by the equations, nor
resolved numerically, can be parameterized within the framework of the
large-scale dynamics. In particular, in a turbulent two-dimensional
flow, enstrophy\footnote{Enstrophy is the ``vortical energy'' of the
flow, defined by $\onehalf\zeta_{\rm rel}^2$, where $\vec\zeta_{\rm
rel} = \zeta_{\rm rel}$~${\bf k} = \nabla\!\times\!{\bf v}$ is the
relative vorticity (parallel to the vertical direction ${\bf k}$).} is
known to cascade to small scales, where three-dimensional effects
eventually become important \citep[e.g.,][]{Pedlosky87}. In the
absence of explicit dissipation in a numerical atmospheric model,
enstrophy accumulates at the smallest resolved scales, in a process
known as spectral blocking, with ever growing numerical errors. It is
thus standard practice in atmospheric science to introduce
hyperdissipation terms in the equations solved to alleviate the
spectral blocking problem \citep[e.g.,][]{Stephenson94}.

In the IGCM, hyperdissipation terms are introduced in the vorticity,
divergence and temperature equations,
\begin{eqnarray} 
Q^{hyp}_{\rm vor} & = & -\nu_{{\rm diss}} (-1)^{{\rm
N_{DEL}}} \nabla ^{2{\rm N_{DEL}}}\, (\zeta - 2 \mu) \\ 
Q^{hyp}_{\rm div} & = & -\nu_{{\rm diss}} (-1)^{{\rm N_{DEL}}} \nabla ^{2{\rm 
N_{DEL}}}\, D \\
Q^{hyp}_T & = & -\nu_{{\rm diss}} (-1)^{{\rm N_{DEL}}} \nabla ^{2{\rm 
N_{DEL}}}\, T_{\rm a}, 
\end{eqnarray}
where $N_{\rm DEL}$ is an integer.  (Note that hyperdissipation is
only applied to the relative vorticity, $\zeta_{\rm rel} = \zeta - 2
\mu$.) Hyper-laplacian operators guarantee that only scales close to
the smallest resolved features are selectively chosen for
diffusion. The case $N_{\rm DEL} =1$, which would correspond to a
regular diffusion operator, is considered to be too widely dissipative
for what are effectively inviscid planetary atmospheres.

It should be noted that hyperdissipation is a numerical tool with
intrinsic limitations. It is not supported by any fundamental theory
and it may well miss some of the ``reverse'' interactions occurring
between small and large horizontal scales in a turbulent atmospheric
flow. There is no rigorous way to chose the magnitude or the order of
hyperdissipation in a given model \citep[e.g.,][]{McVean83} and
different choices have typically been made by different modeling
groups in Earth atmospheric studies \citep[e.g.,][]{Stephenson94}. In
the context of exoplanet atmospheric modeling, with a priori unknown
circulation regimes, hyperdissipation choices should probably be
considered as important free parameters of the models. In keeping with
standard practice \citep{McVean83,Stephenson94}, we use a highly
scale-selective $N_{\rm DEL} =4$ scheme and a dissipation rate
adjusted with horizontal resolution so that structures on the smallest
resolved scales are dissipated in a fraction ($\sim 0.25$ -- $0.01$)
of a planetary day.

\subsection{Numerical Solutions}

The IGCM solves equations~(\ref{eq:vor})--(\ref{eq:hydsta}) with the
semi-implicit pseudo-spectral method \citep{HoskSimm75}. The equations
are solved in spectral space horizontally and with a finite difference
scheme in the vertical. A standard decomposition in spherical
harmonics is used with triangular mode truncation
\citep{Orszag70,Eliassen70,WashPark95}. The pseudo-spectral method is
well adapted to problems with spherical geometry and thus does not
require any special treatment at the poles. The vertical finite
difference scheme is based on the formulation of \citet{SimmBurr81},
which conserves angular momentum and energy exactly. As is usually the
case with this formulation, total atmospheric mass is not conserved by
the vertical scheme but is instead corrected for at each time-step.
Vertical levels are linearly spaced in $\sigma$. In what follows,
model resolutions are reported as T$p$L$q$, which corresponds to
$3p+2$ spectral modes in longitude, half that in latitude, and $q$
vertical levels.  Time integration is performed with a semi-implicit
leapfrog scheme, followed by a Robert-Asselin filter to control time
splitting \citep{HoskSimm75,deFor2000}. A significant advantage of the
semi-implicit method is that it solves gravity wave propagation in an
implicit manner. Larger time steps are thus possible since CFL
stability is determined by advection, rather than the speed of gravity
waves.

Boundary conditions are applied only at the top ($\sigma =0$) and
bottom ($\sigma =1$) of the numerical domain, where $d \sigma / d t
=0$ is imposed. By default, this corresponds to free-slip boundary
conditions for the horizontal flow. All our models are started at
rest, with a temperature profile satisfying everywhere the equilibrium
profile $T^{\rm vert}_{\rm eq}(z)$ in equation~(\ref{eq:teqz}).  A
small amount of noise is introduced to break flow symmetries.

\subsection{Flow Representations}

There are various ways to represent a three-dimensional atmospheric
flow. In the interest of clarity and conciseness, we only show
cylindrical maps and zonally-averaged contour plots. Our temperature
and velocity maps use Miller cylindrical projections, centered on the
equator (and the substellar point, at $[\lambda,\phi]=[0,0]$, in the
case of the hot Jupiter model). In some cases, only a subset of all
velocity vectors are shown, for better rendering. We also present
zonally-averaged contours of zonal velocity, $u$, and temperature,
$T$. These were obtained by performing the longitudinal average
\begin{equation}
[X(t)] \equiv \left[ X(\sigma,\phi, t) \right] = \frac{1}{2 \pi}
\int_0^{2 \pi} X(\sigma,\lambda, \phi, t) \ d \lambda,
\end{equation}
where $X = u$ or $T$.

\section{Earth-Like Model} \label{sec:Earth}

Our Earth-like model comes as a default implementation of the IGCM
solver. It has many similarities with the control run described by
\citet{JamesGray86}. To capture in a simple way Earth's annual mean
conditions, a meridional gradient of relaxation temperatures is
imposed, such that
\begin{equation}
\Delta T_\theta (\lambda,\phi) = \Delta T_\theta (\phi) =( \frac{1}{3} - \sin^2 \phi ) \times \Delta T_{\rm EP},
\end{equation}
where the equator-pole temperature difference is set to $\Delta T_{\rm
EP} =60$~K.  Parameters for the vertical relaxation profile, $T^{\rm
vert}_{\rm eq}(z)$ in equation~(\ref{eq:teqz}), are chosen to capture
approximately Earth's typical radiative-convective conditions. For
simplicity, a single value for the radiative relaxation time,
$\tau_{\rm rad} =15$~planet days, is adopted throughout the
atmosphere. Rayleigh drag with a friction time $\tau_{\rm
fric}=1$~planet day is applied to the lowest model layer only. There
is no account of seasonal variations or topography in the
model. Stratospheric conditions are rather poorly captured. The
complete list of parameters of our Earth-like model with T42L15
resolution is provided in Table~\ref{tab:one}.

This model also share strong similarities with the classic benchmark
calculation of \citet{HeldSuar94} for dynamical core validations. A
significant difference with the Held-Suarez benchmark is that the
variations of $\tau_{\rm rad}$ and $\tau_{\rm fric}$ values with
location in the atmosphere in that model are reduced to single values
in the simpler model presented here. Our Earth-like model may thus be
considered as a crude version of the Held-Suarez benchmark, with a
reduced number of free parameters.

The main features of Earth's general atmospheric circulation and the
ability of idealized models such as the Held-Suarez benchmark to
reproduce them with reasonable accuracy are well known
\citep[e.g.,][]{HeldSuar94}. The key features of the tropospheric
circulation, at $p \gsim 0.1$~bar ($\sigma \gsim 0.1$) levels, are:
(i) an axisymmetric meridional circulation via Hadley cells extended
from the equator to approximately $\pm 30 \degr$ in latitude, which
efficiently reduce meridional temperature gradients in that region,
(ii) strong baroclinic activity at mid-latitudes, with typically $\sim
6$ large scale baroclinic eddies spread in longitude along a ``storm
track,'' (iii) a zonal circulation at the surface which is
characterized by easterly (westward) trade winds in subtropical
latitudes, westerlies (eastward wind) in mid-latitudes and weak
easterlies near the poles, (iv) a westerly component of the zonal wind
increasing with height at all latitudes, until localized wind maxima
known as the jet streams are reached at the $\sim 0.2$~bar ($\sigma
\sim 0.2$) level, with peak winds at $ \pm 40$--$50\degr$ latitude.

\begin{figure}[ht]
\begin{center}
\vspace{-3.5cm}
\includegraphics[scale=0.85]{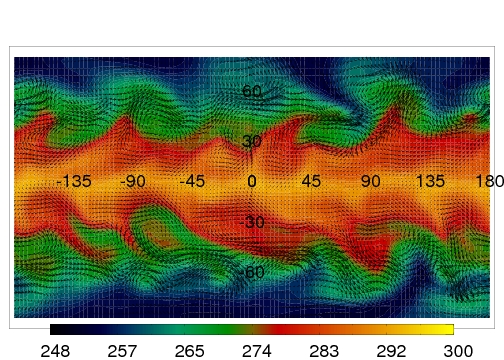}
\includegraphics[scale=0.85]{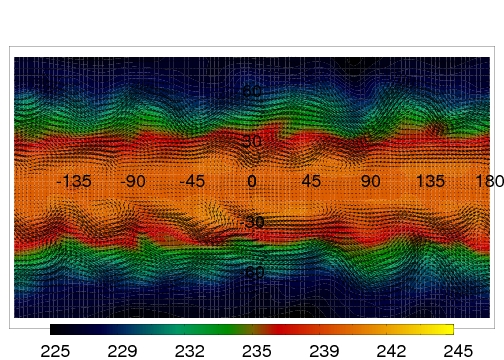}
  \caption{Cylindrical maps of temperature and velocity at planet day
  150 in the Earth-like model with T42L15 resolution. The color scale
  shows temperatures in K.  The top panel illustrates the strong
  baroclinic activity present at mid-latitudes in the bottom model
  layer, at the $\sigma =0.97$ level. The bottom panel illustrates the
  reduced level of baroclinic activity and the formation of two
  mid-latitudinal jet streams in the upper troposphere, shown here at
  the $\sigma =0.37$ level.  }\label{fig:one}
\end{center}
\end{figure}

\begin{figure}[ht]
\begin{center}
\vspace{-2.5cm}
\includegraphics[scale=0.85]{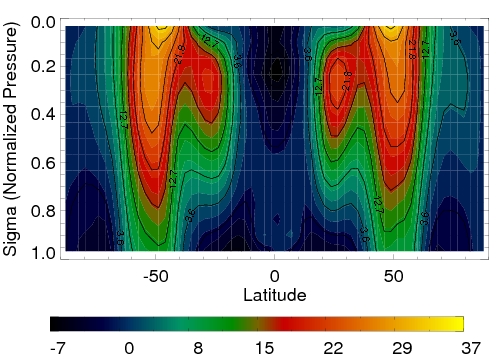}
\includegraphics[scale=0.85]{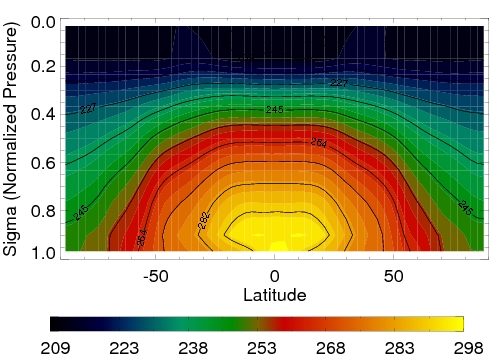}
  \caption{Zonally-averaged contours of zonal wind speed ([U] in m/s,
  top panel) and temperature ([T] in K, bottom panel) for the same
  Earth-like model as shown in Figure~\ref{fig:one}, at planet day
  150. Sigma is the pressure normalized to the bottom value (1
  bar). These results are in qualitative agreement with the
  Held-Suarez benchmark, even though quantitative differences exist
  (see text for details).}\label{fig:two}
\end{center}
\end{figure}

Figures~\ref{fig:one} and~\ref{fig:two} illustrate how our simple
Earth-like model qualitatively reproduces this general circulation
regime. While time averages over hundreds of planet days are
traditionally used to characterize the flow in the Held-Suarez
benchmark \citep{HeldSuar94}, we have chosen to present snapshots of
the atmospheric flow for simplicity and to facilitate comparisons with
the hot Jupiter model presented below. Figure~\ref{fig:one} shows
temperature and velocity maps at planet day 150 in the Earth-like
model. The top panel exemplifies the strong baroclinic activity that
characterizes mid-latitudes, shown here in the bottom model layer at
the $\sigma =0.97$ ($\simeq 0.97$ bar) level. The bottom panel
exemplifies the reduced baroclinic activity and the formation of jet
streams that characterize the upper troposphere, shown here at the
$\sigma =0.37$ ($ \simeq 0.37$ bar) level.

Figure~\ref{fig:two} shows zonally-averaged contours of zonal wind
speeds ($[u]$ in m~s$^{-1}$; top panel) and temperature ($[T]$ in K;
bottom panel) for the same Earth-like flow at planet day 150 as shown
in Figure~\ref{fig:one}. While the $[u]$ contours reproduce the main
qualitative features of the Held-Suarez benchmark calculation
\citep[see Fig.~2 in][]{HeldSuar94}, quantitative discrepancies
emerge, most notably in the extremas of wind speeds and in the
detailed shape of the zonal wind structure (with jet stream cores
incorrectly pushed against the top layer in our model). We attribute
these differences to the simpler nature of Newtonian forcing and
Rayleigh drag in our model, different relaxation profiles and our
focus on a flow snapshot rather long-term averages. Similarly, a
comparison of the $[T]$ contours shown in the lower panel of
Figure~\ref{fig:two} with the corresponding Figure~1c in
\citet{HeldSuar94} reveals broad qualitative agreement (e.g.,
flattened equatorial contours) but also quantitative
discrepancies. These discrepancies can be partly attributed to the
different vertical profiles of relaxation temperature adopted here and
in \citet{HeldSuar94}.

Rather than focusing on a strict reproduction of the Held-Suarez
benchmark results, which is of limited interest for a well-tested code
like the IGCM solver \citep[see, e.g., the climatology
of][]{deFor2000}, our simple Earth-like model may offer interesting
insight into important issues of model parameterizations and target
accuracies for exoplanet atmospheric circulation studies. By
comparison with a somewhat higher complexity model like the
Held-Suarez benchmark calculation, it provides a measure of the
ability of strongly parameterized models to successfully reproduce the
main qualitative features of an atmospheric circulation regime like
that on Earth. We note, however, that in both our model and the
Held-Suarez benchmark, parameters were adjusted a posteriori to
reproduce a known circulation regime. By contrast, circulation regimes
are a priori unknown on exoplanets and it may be difficult to
determine from first principles the temperature relaxation profiles
and relaxation times needed to adequately drive or drag the flow on a
remote planet. On the other hand, it could also be that qualitative
agreement at a level comparable to that achieved by our simple
Earth-like model turns out to be sufficient to interpret with
confidence typical remote astronomical observations of exoplanets. The
issue of target accuracies for reliable exoplanet data interpretation
has received little attention until now. We will simply note here
that, in addition to validations and inter-comparisons,
parameter-space explorations with simple atmospheric circulation
models may be important ingredients of an effective strategy to
address this data interpretation challenge.

\section{Shallow Hot Jupiter Model} \label{sec:HJ}

To capture the permanent day--night forcing conditions present on a
tidally-locked hot Jupiter, our IGCM solver has been modified to
permit horizontal gradients of relaxation temperatures of the form
\begin{equation}
\Delta T_\theta (\lambda,\phi) = \cos \lambda \cos \phi \times \Delta
T_{\rm EP}, 
\end{equation}
which places the substellar point at $(\lambda, \phi) = (0,0)$.  In
the specific hot Jupiter model presented here, the equator-pole
temperature difference is set to $\Delta T_{\rm EP} =300$~K, which
corresponds to a full day-night temperature difference of $2 \Delta
T_{\rm EP} =600$~K. The amplitude of this day-night temperature
forcing, which is an important free-parameter of our model, is
comparable to the corresponding forcing amplitude at the $\sim 1$~bar
level in the circulation model of \citet{C&S05}. A relatively steep,
linear dependence of the equilibrium relaxation temperature with the
cosine of the angle away from the substellar point has been adopted
because it accounts for the extra atmospheric depth crossed by
radiation at inclined angles and is broadly consistent with detailed
radiative transfer calculations \citep[e.g.,][]{Showmanetal08}.

Parameters for the vertical relaxation profile, $T^{\rm vert}_{\rm
eq}(z)$ in equation~(\ref{eq:teqz}), are chosen to match approximately
the profile of \citet{Iro05} in the $\sim 0.1$-$1$~bar region for
\HDblah, with a constant lapse rate $\Gamma_{\rm trop} = 2 \times
10^{-4}$~K~m$^{-1}$ and no stratosphere. A significant feature of our
vertical relaxation profile is that the day-night temperature
differential asymptots to zero in the uppermost modeled layers, like
it does in the Earth-like model (see
Eqs.~[\ref{eq:Teq}--\ref{eq:Teq2}]). While this choice facilitates a
direct comparison of circulation regimes between the Earth-like model
(with meridional forcing) and the hot Jupiter model (with hemispheric
forcing), it may also be a poor assumption for a hot Jupiter
atmosphere. More realistically, the day-night temperature differential
would extend to layers higher up in the atmosphere (beyond those
modeled), with the possible existence of a stratosphere depending on
the presence of an absorbing compound such as TiO/VO or Sulfur
\citep{Hubeny03,Burrows07,Fortney2008,Spiegel09,Zahnle09}.

We adopt a single value for the radiative relaxation time, $\tau_{\rm
rad} =0.5$~planet day $\simeq 1.5 \times 10^5$~s, to match the
radiative timescale at 1 bar from \citet{Iro05}.  While the radiative
times are expected to vary substantially with depth in hot Jupiter
atmospheres, a single $\tau_{\rm rad}$ value, like in our Earth-like
model, is the simplest acceptable form of radiative forcing in a
shallow atmospheric model such as ours. No Rayleigh drag is
implemented. All the other parameters are chosen appropriately for the
hot Jupiter \HDblah. The complete list of parameters of this idealized
hot Jupiter model with T42L15 resolution is provided in
Table~\ref{tab:one}. We emphasize that the model is quite shallow in
the sense that only 1--2 vertical levels are present above the $\sigma
=0.1$ level (given the linear-$\sigma$ grid and vertical resolutions
used) and that absolutely no account is made of deeper atmospheric
layers present below the 1~bar pressure level (which bounds our model
at $\sigma =1$). Despite its great simplicity, a clear advantage of
this shallow hot Jupiter model is that it is a direct extension of our
Earth-like model setup to the case of a hot Jupiter atmospheric flow
and thus permits straightforward comparisons between the two simulated
circulation regimes.

\begin{figure}[ht]
\begin{center}
\vspace{-3.5cm}
\includegraphics[scale=0.85]{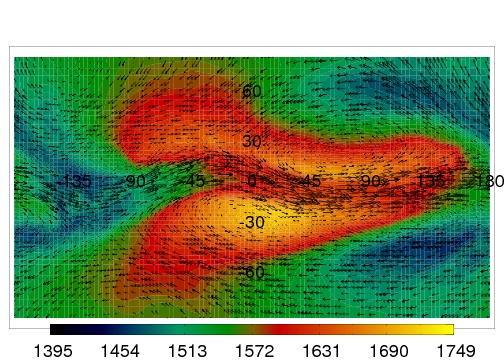}
\includegraphics[scale=0.85]{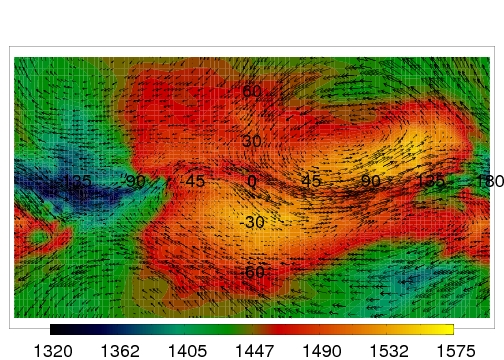}
  \caption{Cylindrical maps of temperature and velocity at planet day
  100 in the shallow hot Jupiter model with T42L15 resolution. The
  color scale shows temperatures in K. Flows in the top ($\sigma =0.7$
  level) and bottom ($\sigma = 0.37$ level) panels share strong
  similarities. This vertical flow alignment and the lack of clear
  baroclinic activity is consistent with a barotropic flow
  regime. Note the presence of large circumpolar vortices.  Advection
  of heat away from the central sub-stellar region occurs both
  westward (at the equator) and eastward (at
  mid-latitudes).}\label{fig:three}
\end{center}
\end{figure}

Figure~\ref{fig:three} shows temperature and velocity maps at planet
day 100 in the hot Jupiter model, at the $\sigma =0.7$ (top panel) and
$0.37$ (bottom panel) levels. The temperature fields and particularly
the velocity fields share strong similarities at these two
levels. This vertical flow alignment, which persists throughout the
various modeled layers, together with the lack of any identifiable
baroclinic eddies, is characteristic of a barotropic flow regime
\citep[e.g.,][]{Cho2003,Menou2003,Cho2006}. The flow is characterized
by a zonally-perturbed superrotating equatorial wind, flanked by
dynamically active vortices, counter-jets at mid-latitudes and the
presence of large scale polar vortices. Advection of heat away from
the substellar point, at $(\lambda, \phi) = (0,0)$, occurs both
eastward in the equatorial regions and westward in mid-latitudes,
where the counter-jets are present.

\begin{figure}[ht]
\begin{center}
\plotone{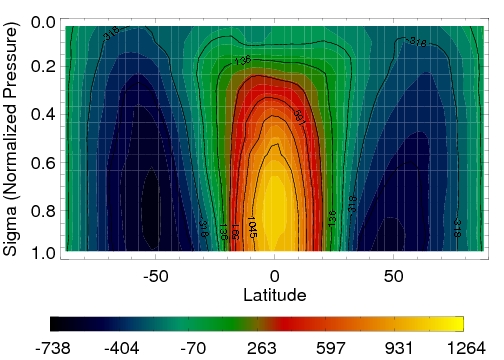}
  \caption{Zonally-averaged contours of zonal wind speed ([U] in m/s)
  for the same shallow hot Jupiter model as shown in
  Figure~\ref{fig:three}, at planet day 100. Sigma is the pressure
  normalized to the bottom value (1 bar). Over the entire vertical
  extent of the modeled region, the flow is characterized by a
  super-rotating equatorial wind and slower counterjet at
  mid-latitudes. Zonal-average wind speeds are well below
  corresponding sound speeds (which are $\gsim 2500$
  m/s).}\label{fig:four}
\end{center}
\end{figure}

Figure~\ref{fig:four} offers another view of the circulation regime in
this shallow hot Jupiter model with zonal averages of the zonal wind
velocity ($[u]$) at planet day 100. Over nearly the entire vertical
extent of the model layer, the flow exhibits the broad super-rotating
(eastward) equatorial wind and slower westward counter-jets at
mid-latitudes. Maximum wind speeds are $\lsim 1800$ m~s$^{-1}$ and
zonal averages are $\lsim 1300$ m~s$^{-1}$. These values are well
below corresponding sound speeds, which range from $2.4$ to $3.1$
km~s$^{-1}$ from top to bottom of the modeled region.

\begin{figure}[ht]
\begin{center}
\vspace{-1cm}
\includegraphics[scale=0.7]{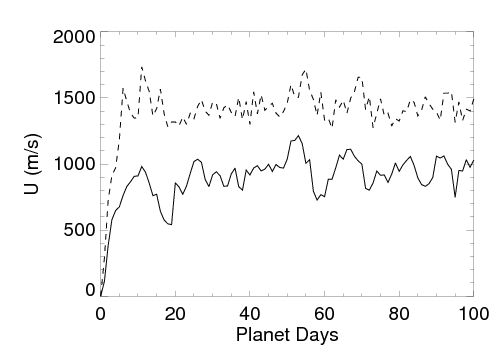}
\includegraphics[scale=0.7]{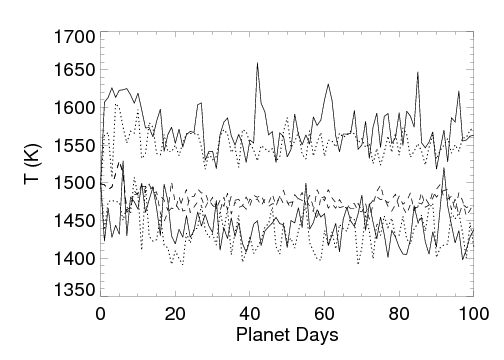}
  \caption{Time evolution of representative velocities and
  temperatures in the same shallow hot Jupiter model as shown in
  Figs~\ref{fig:three} and~\ref{fig:four}. All quantities are shown at
  specific locations on the $\sigma =0.5$ level. Flow steady-state is
  achieved after $\sim 20$ days.  Top panel: zonally-averaged (solid
  line) and maximum (dashed line) value of the zonal velocity $u$
  along the equator.  Bottom panel: temperatures at the sub- and
  antistellar points (top and bottom solid lines, respectively), at
  the equatorial east and west limbs (top and bottom dotted lines,
  respectively) and at the north and south poles (two dashed lines).
  Asymptotic wind speeds are subsonic. Temperatures are far from being
  horizontally homogenized.  }\label{fig:five}
\end{center}
\end{figure}

Various diagnostics can be used to verify that the flow has reached a
stationary state with respect to the imposed forcing. We have
performed extended runs for up to several hundred planetary days and
have found stationary conditions for the shallow hot Jupiter model
under consideration. To illustrate this, Figure~\ref{fig:five} shows
the time evolution over 100 planet days of representative velocities
and temperatures at various horizontal locations on the $\sigma = 0.5$
model level. The zonal average and maximum values of the zonal
velocity $u$ along the equator are shown in the top panel as solid and
dashed lines, respectively. After a rapid acceleration phase lasting
$\sim 5$--$10$ planet days, a flow stationary state is reached at
planet day $\sim 20$, with significant fluctuations. Throughout this
evolution, flow velocities remain subsonic.  In the bottom panel of
Figure~\ref{fig:five}, the corresponding evolution of temperatures is
shown at the sub- and antistellar points (top and bottom solid lines,
respectively), at the east and west equatorial limbs (top and bottom
dotted lines, respectively) and at the north and south poles (two
dashed lines). Despite eastward heat advection at the equator, which
results in comparable temperatures at the east equatorial limb and the
substellar point, temperatures are far from being horizontally
homogeneous. Note in particular that temperatures around the planetary
limb (dashed and dotted lines) represent a diverse range of physical
conditions even on a fixed ($\sigma = 0.5$) pressure level (see also
Fig.~\ref{fig:three}).

\begin{figure}[ht]
\begin{center}
\vspace{-3.5cm}
\includegraphics[scale=0.85]{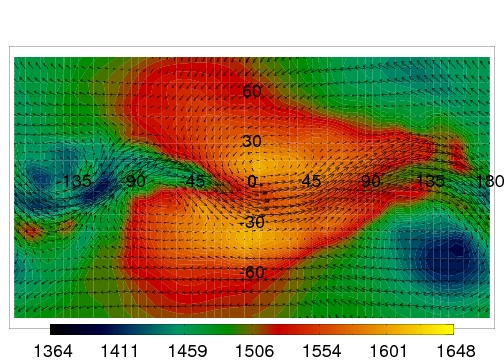}
\includegraphics[scale=0.85]{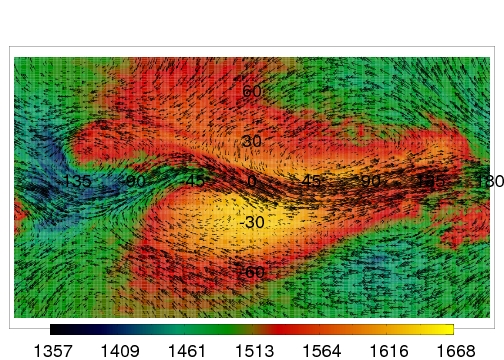}
\caption{A test of numerical convergence. Cylindrical maps of
  temperature and velocity at planet day 100 in shallow hot Jupiter
  models with T21L5 resolution ($\sigma = 0.5$ level, top panel) and
  T170L20 resolution ($\sigma = 0.52$ level, bottom panel). The color
  scale shows temperatures in K. The overall similarity of these two
  maps indicates good overall convergence. Even the T21L5 flow shares
  many of the global attributes of higher resolution simulated flows
  (see text for details). }\label{fig:six}
\end{center}
\end{figure}

We have found that these results are broadly confirmed at different,
and in particular higher, model resolutions. Figure~\ref{fig:six}
presents a specific test of numerical convergence for our
results. Temperature and velocity maps at planet day 100 are shown for
the same shallow hot Jupiter model as before, except that reduced and
enhanced numerical resolutions were used, both horizontally and
vertically. The top panel shows a map at the $\sigma =0.5$ level in a
T21L5 model. The bottom panel shows a corresponding map, at the
$\sigma =0.52$ level, in a T170L20 model. Values of the
hyperdissipation coefficient $\nu_{\rm diss}$ were adjusted to $7.3
\times 10^{49}$ and $4.7 \times 10^{43}$~m$^8$~s$^{-1}$ in these T21L5
and T170L20 models, respectively. The overall similarity of these
temperature and velocity maps, as well as other flow attributes (e.g.,
zonal wind contours), indicates that good convergence is already
achieved around T31L10 to T42L15 resolution. Even the T21L5 flow
shares many of the global attributes of higher resolution simulated
flows.

As mentioned earlier, the zonally-perturbed equatorial wind and its
flank vortices are dynamical features of the simulated
flow. Figure~\ref{fig:seven} illustrates the flow unsteadiness with
two successive temperature and velocity maps at planet days 97 and 98
in a T85L20 version of our shallow hot Jupiter model (with $\nu_{\rm
diss}$ adjusted to $5.9 \times 10^{45}$~m$^8$~s$^{-1}$). The maps are
shown at the $\sigma =0.52$ level. While this particular example was
chosen to exhibit clear temperature and flow field variability over
one planet day, dynamical variability is a general property of the
simulated flow (see fluctuations in Fig.~\ref{fig:five}). Nevertheless,
we find that the displaced anticyclonic polar vortices that emerge in
this hot Jupiter model do not experience systematic longitudinal
translations, nor are they close to geostrophic balance, like the
cyclonic circumpolar vortices discussed by \citet{Cho2003,Cho2006}.
Instead, the polar vortices in the present model appear to be strongly
tied to the imposed day-night forcing and show only limited excursions
away from their preferred night-side location, somewhat eastward of
the anti-stellar point (see Figs.~\ref{fig:three}, \ref{fig:six}
and~\ref{fig:seven}).

The wind acceleration episode apparent in Figure~\ref{fig:five} must
be important in determining the nature of the stationary regime
eventually achieved by the forced flow. We have found evidence that
barotropic instabilities play a role in shaping the dynamical nature
of this flow. During the first few planet days in our hot Jupiter
model, we observe the acceleration of a zonal eastward wind in the
equatorial regions and westward counter-jets at mid-latitudes. This
situation is very reminiscent of the counter-acceleration caused by
meridional Rossby wave transport that occurs in the idealized
momentum-forced flow discussed by \citet{Cho2006} in the context of
the equivalent-barotropic formulation (see their
Fig.~17). Figure~\ref{fig:eight} shows a temperature and velocity map
at planet day 5 and level $ \sigma =0.9$ in our T42L15 hot Jupiter
model. In contrast with previous maps, this one zooms in a specific
region along the equator, restricted $+20$ to $+160 \degr$ in
longitude and $\pm 40 \degr$ in latitude. In this region, the flow
exhibits strong horizontal (zonal) shear, as well as small scale
velocity and temperature disturbances along the leading edge of the
equatorial jet. Subsequently, we observe a rapid thinning of the jet,
a breaking of equatorial symmetry by planet day $6$-$7$ and the
emergence of a broader, wavy equatorial wind, as shown for instance in
Figure~\ref{fig:three}.

Figure~\ref{fig:nine} suggests that horizontal shear (= barotropic)
instabilities are important dynamical ingredients of the sequential
flow evolution we have just described. The latitudinal profile of
zonally-averaged zonal wind velocity, $[u]$, is shown in the top
panel, while its second-order meridional derivative, $d[u]/dy^2$, is
shown in the bottom panel, for the same planet day 5 and $ \sigma
=0.9$ level flow as in Fig.~\ref{fig:eight}. Although necessary and
sufficient conditions for the development of barotropic (horizontal
shear) instabilities are generally not known, the Rayleigh-Kuo
inflexion point criterion provides a useful necessary condition, which
accounts for the stabilizing influence of the planetary vorticity
\citep{Kuo1949,Vallis2006}. The instability condition is met when
$d[u]/dy^2$ exceeds the planetary parameter $\beta \equiv df / d \phi
= 2 \Omega_p \cos \phi / R_p$ (latitudinal gradient of the Coriolis
parameter), which is shown as a dotted line in the bottom panel of
Figure~\ref{fig:nine}. While the moderate violations of the
Rayleigh-Kuo criterion at mid-latitudes and beyond may not be very
meaningful\footnote{Away from the equator, the flow is not as strongly
zonal as in the equatorial regions shown in
Fig.~\ref{fig:eight}. Applying a zonal instability criterion in these
regions may thus be of limited value. }, the strong violations on each
side of the equatorial wind, at $\sim \pm 20 \degr$ latitude, are
consistent with the substantial zonal shear present there and the
associated small scale disturbances shown in Fig.~\ref{fig:eight}.

We note that the significance of barotropic instabilities operating in
the flow, as suggested by Figures.~\ref{fig:eight} and~\ref{fig:nine},
is that they could play an important dynamical role by tapping the
free energy available in the horizontal shear flow and thus possibly
limit the asymptotic speeds of winds in our model. While wind
acceleration followed by saturation in the first $\sim 10$ days, as
shown in Figure~\ref{fig:five}, appears to be broadly consistent with
this notion, additional flow diagnostics beyond the scope of the
present study would be needed to establish more confidently this
possibility.

\section{Discussion and Conclusion}

In this work, we have presented a simple Earth-like general
circulation model based on the IGCM dynamical core. We used this model
to contrast the baroclinic circulation regime of the Earth's lower
atmosphere with the more barotropic circulation regime that emerges
from a straightforward extension of the model to the atmospheric flow
on a hot Jupiter.  The distinction between these two major (barotropic
and baroclinic) regimes of atmospheric circulation is an important one
in ``geophysical'' fluid dynamics. For instance, both regimes are
relevant to the Earth's atmosphere and critical to our understanding
of its general circulation, with a baroclinic (lower-level)
troposphere and a barotropic (higher-level) stratosphere.

Various factors contribute to the degree of
baroclinicity/barotropicity of an atmospheric flow. The more
stably-stratified an atmosphere is (i.e. the more ``radiative'' it is,
as opposed to convective, to use the language of stellar physics), the
larger its external Rossby deformation radius is, the weaker
baroclinic instability growth is and thus the more barotropic the
circulation regime will be (e.g. Pedlosky 1987; Cho et al. 2008).  The
strong external irradiation experienced by hot Jupiter atmospheres
creates rather strongly-stratified temperature profiles in their
photospheric regions (e.g. Seager \& Sasselov 1998; Sudarsky et
al. 2000; Iro et al. 2005; Barman et al. 2005; Fortney \& Marley
2007). This relative vertical stability, together with slow
(synchronized) rotation and high atmospheric temperatures, leads to
large external Rossby deformation radii
\citep{Showman2002,Cho2003,Menou2003,Cho2006} and favors a barotropic
circulation regime (with vertically aligned horizontal motions in the
various atmospheric layers). The lack of surface drag on the
atmospheric flow, for otherwise identical forcing conditions, also
favors a barotropic regime as horizontal shear tends to inhibit the
development of baroclinic instabilities
\citep[e.g.,][]{JamesGray86,James1987,Robinson1997} For all these
reasons, a single-layer, vertically-integrated barotropic treatment of
horizontal motions in hot Jupiter atmospheres may be justified
\citep{Cho2003,Cho2006,Menou2003,Salby1989}.

Since the distinction between barotropic and baroclinic regimes
depends on the degree of atmospheric vertical stratification, which is
known for the Earth but {\it a priori} unknown for remote exoplanets,
the results from our shallow hot Jupiter model should only be
interpreted as suggestive that this regime is relevant to hot Jupiter
atmospheric flows. A more systematic exploration of circulation
regimes on hot Jupiters will be needed to address this issue more
thoroughly.  The shallow hot Jupiter model presented here is rather
specific and idealized in a number of important ways. For instance,
adopted values for the profile of relaxation temperatures and the
radiative relaxation time are rather arbitrary.  The presence of
deeper atmospheric layers and their interaction with the modeled
layers has also been ignored in this shallow model.

\begin{figure}[ht]
\begin{center}
\vspace{-3.5cm}
\includegraphics[scale=0.85]{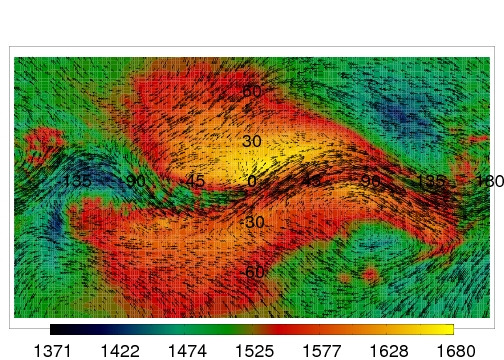}
\includegraphics[scale=0.85]{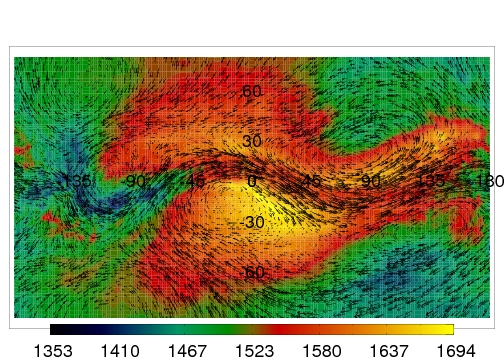}
  \caption{The flow unsteadiness is illustrated by comparing two
  temperature and velocity cylindrical maps at planet days 97 and 98
  in the shallow hot Jupiter models with T85L20 resolution (at the
  $\sigma = 0.52$ level). The color scale shows temperatures in
  K. Both the temperature and flow fields exhibit significant
  variability over one planet day.}\label{fig:seven}
\end{center}
\end{figure}

\begin{figure}[ht]
\begin{center}
\plotone{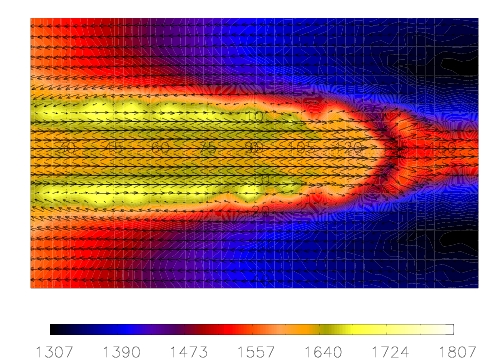}
  \caption{Cylindrical map of temperature and velocity at planet day 5
  in the shallow hot Jupiter model with T42L15 resolution, at the
  $\sigma = 0.9$ level. The color scale shows temperatures in K. The
  map zooms in a specific equatorial region ($\pm 40 \degr$ in
  latitude, $+20$ to $+160 \degr$ in longitude), where the
  accelerating equatorial flow exhibits strong horizontal shear and
  small scale disturbances. }\label{fig:eight}
\end{center}
\end{figure}

\begin{figure}[ht]
\begin{center}
\vspace{-1cm}
\includegraphics[scale=0.7]{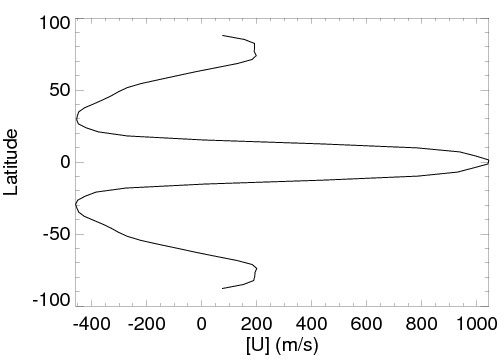}
\includegraphics[scale=0.7]{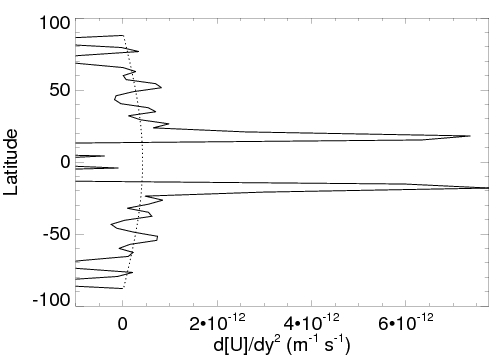}
  \caption{Zonally-averaged zonal wind ($[u]$, top panel) and its
  second derivative with respect to latitudinal length ($d[u]/dy^2$,
  bottom panel) in the same shallow hot Jupiter model with T42L15
  resolution as shown in Fig.~\ref{fig:eight} (planet day 5 and
  $\sigma = 0.9$ level). The Rayleigh-Kuo necessary condition for
  barotropic instability is satisfied when $d[u]/dy^2$ exceeds the
  planetary $\beta$ parameter (shown as a dotted line in the bottom
  panel). The strong violation at $ \pm 20 \degr$ latitude is
  consistent with the substantial horizontal shear and associated
  small scale disturbances seen in
  Fig.~\ref{fig:eight}.}\label{fig:nine}
\end{center}
\end{figure}

Nevertheless, the model may capture important dynamical features of
the atmospheric flow on hot Jupiters. In particular, the simulated
flow has a number of similarities with comparable results reported in
the literature, together with noticeable differences. As we have
already emphasized, our simulated flow is characterized by a broad,
zonally-perturbed superrotating equatorial wind, large scale polar
vortices, unsteadiness and subsonic wind speeds. The emergence of a
super-rotating (eastward equatorial) wind in this
hemispherically-forced flow is consistent with the results of
\citet{Showman2002,C&S05,Showmanetal08,Dobbs-Dixon2008,Langton2008}. The
broad width of this equatorial wind and the presence of counter-jets
at midlatitudes is also consistent with the specific results of
\citet{Showmanetal08} and \citet{Cho2003,Cho2006}. As discussed in
detail by \citet{Cho2006}, however, the typically opposite (westward)
direction of the equatorial wind that emerges in equivalent-barotropic
simulations may point to some limitation of that approach. On the
other hand, the strong zonal disturbances of the equatorial wind in
our model is very reminiscent of a similar pattern of large-amplitude
Rossby waves discussed by Cho et al. (2003; 2008; see also Langton \&
Laughlin 2007). This is qualitatively different from the
zonally-symmetric equatorial flow reported by
\citet{Showman2002,C&S05,Showmanetal08}. As Fig.~\ref{fig:seven}
illustrates, these zonal disturbances of the equatorial wind are
closely related to the flow unsteadiness observed in our model
\citep[see also][]{Langton2008}.

While our shallow hot Jupiter flow exhibits large-scale polar
vortices, their dynamical nature is quite distinct from that of the
circumpolar vortices discussed by \citet{Cho2003,Cho2006}.  The polar
vortices shown in Fig.~\ref{fig:seven}, for instance, are
anti-cyclonic, not geostrophically-balanced and consistently located
on the planetary night-side, somewhat eastward of the anti-stellar
point. This is in contrast with the geostrophically-balanced, cyclonic
polar vortices that exhibit systematic longitudinal translations in
the equivalent-barotropic flows described by
\citet{Cho2003,Cho2006}. We interpret this difference as being due to
the forcing-dominated nature of polar vortices in our hot Jupiter
model, which starts at rest, rather than a dynamical origin like in
the flows of \citet{Cho2003,Cho2006}, where polar vortices emerge from
turbulent initial conditions subject to an energy cascade to large
scales under the constraint of potential vorticity conservation (see
\citealt{Cho2003,Cho2006} for a discussion; see also
\citealt{Langton2007}). We note that, even though our simulated flow
is clearly unsteady (see Fig.~\ref{fig:seven}), a possible consequence
of the different nature of polar vortices in the present shallow hot
Jupiter model could be a reduced level of disk-integrated variability,
from less dynamically active polar vortices, by comparison to what
equivalent-barotropic results have indicated so far
\citep{Cho2003,Menou2003,Cho2006,Rauscher2007,Rauscher2008}.

An important difference between our results and several comparable
studies reported in the literature is the consistently subsonic value
of wind speeds in our shallow hot Jupiter model.  In this respect, our
results stand out by comparison with those of
\citet{Showman2002,C&S05,Showmanetal08,Dobbs-Dixon2008,Langton2008}. It
is presently unclear what is the origin of this fundamental
discrepancy. We simply note here that one element of answer might be
the emergence of barotropic (horizontal shear) instabilities in our
shallow model, which appear to result from the acceleration of the
equatorial wind and its flank counterjets and could possibly limit the
asymptotic wind speeds in our model. More work, including model
inter-comparisons, is needed to clarify this point.

The magnitude of wind speeds on hot Jupiters is an open
problem. Unlike the atmospheres of terrestrial planets, giant planet
atmospheres lack the large-scale sink of energy and momentum that is
associated with friction on a solid surface.\footnote{On the Earth and
in our simple Earth-like model, for instance, the atmosphere reaches a
global state of momentum balance with the bulk planetary rotation
through surface drag, with positive and negative contributions
depending on the easterly or westerly nature of surface winds
\citep[e.g.,][]{Holton1992,JamesGray86}. } As a result, the only
sources and sinks of energy and momentum in a hot Jupiter flow as
simulated here are the Newtonian relaxation, which represents
large-scale sources and sinks of radiation, and the hyperdissipation,
which operates on small scales. The absence of a clearly identified
large-scale sink of energy (ground friction) makes giant planet
atmospheres possibly more difficult to understand than their
terrestrial counterparts.  Indeed, dissipation in the flow, which
ultimately determines asymptotic wind speeds, is then more strongly
dependent on the flow itself, via its large-scale coupling to
radiation and its effective turbulent dissipation on small
scales. \citet{Goodman09} has recently suggested that internal
friction between atmospheric layers could play an important role in
the hot Jupiter context.

The large variety of flow behaviors found so far in distinct hot
Jupiter studies suggests that, in addition to model validations such
as our simple Earth-like model, inter-comparisons of hydrodynamic
solvers on identical, well-defined atmospheric circulation problems
may be necessary to build a solid understanding of these new
circulation regimes. The shallow hot Jupiter model presented here
could be used as a first step in this direction.

\section*{Acknowledgments}

This work was supported by NASA contract NNG06GF55G. It has benefited
from numerous scientific exchanges with James Cho.

\clearpage

\begin{deluxetable}{lcc} 
\tablewidth{0pt}
\tablecaption{Model Parameters}
\tablehead{
\colhead{Parameters}  & \multicolumn{2}{c}{Model} \\
\colhead{~}  & \colhead{Earth-like}  & \colhead{Hot Jupiter}  
}
\startdata
$g$ (gravitational acceleration [m s$^{-2}$])&  9.81   &  8     \\
$\Omega_p$  (planetary rotation rate [rad s$^{-1}$]) &  $7.292 \times 10^{-5}$   &  $2.1 \times 10^{-5}$      \\
$R_p$  (planetary radius  [m]) &  $6.371 \times 10^6$   &  $10^8$      \\
${\cal R}$  (perfect gas constant [MKS]) &  287   &  3779      \\
$\kappa$  ($={\cal R}/c_p$ [MKS]) &  0.286   &  0.286      \\ 
\\ \hline \\
Resolution (T--horizontal; L--vertical)& T42L15 & T42L15 \\ 
$2{\rm N_{DEL}}$  (hyperdissipation order) & 8   &  8      \\
$\nu_{\rm diss}$  (hyperdissipation value [m$^8$ s$^{-1}$]) &  $1.18 \times 10^{37}$   &  $6.28 \times 10^{47}$    \\
\\ \hline\\
$\tau_{\rm fric}$  (Rayleigh friction time -- bottom layer [planet days]) &  1   &  $\infty$      \\
$\tau_{\rm rad}$  (Newtonian relaxation time -- all layers [planet days]) &  15   &  0.5      \\
$\Delta T_{\rm EP}$  (equator-pole  difference for $T_{\rm eq}$ [K]) &  60   &  300      \\
$\Gamma_{\rm trop}$  (tropospheric lapse rate for $T_{\rm eq}$ [K m$^{-1}$]) &  $6.5 \times 10^{-3}$  &  $2 \times 10^{-4}$      \\
$T_{\rm surf}$ (base value for $T_{\rm eq}$ [K]) & 288 & 1600 \\
$z_{\rm stra}$ (height of tropopause for $T_{\rm eq}$ [m]) & $1.2 \times 10^4$ & $2 \times 10^6$ \\
$\delta T_{\rm stra}$ (tropopause temperature increment for $T_{\rm eq}$ [K]) & 2 & 10 \\
\enddata
\label{tab:one}
\end{deluxetable}

\clearpage

\end{document}